# Machine Learning Based Prediction of Proton Conductivity in Metal-Organic Frameworks

*Seunghee Han[1], Byoung Gwan Lee[2], Dae Woon Lim[2], and Jihan Kim[1*]*

1 Department of Chemical and Biomolecular Engineering, Korea Advanced Institute of Science and Technology, Daejeon 34141, Republic of Korea.

2 Department of Chemistry and Medical Chemistry, Yonsei University, Wonju, Gangwondo 26493, Republic of Korea

*Corresponding author: jihankim@kaist.ac.kr

# ABSTRACT

Recently, metal-organic frameworks (MOFs) have demonstrated their potential as solid-state electrolytes in proton exchange membrane fuel cells. However, the number of MOFs reported to exhibit proton conductivity remains limited, and the mechanisms underlying this phenomenon are not fully elucidated, complicating the design of proton-conductive MOFs. In response, we developed a comprehensive database of proton-conductive MOFs and applied machine learning techniques to predict their proton conductivity. Our approach included the construction of both descriptor-based and transformer-based models. Notably, the transformer-based transfer learning (Freeze) model performed the best with a mean absolute error (MAE) of 0.91, suggesting that the proton conductivity of MOFs can be estimated within one order of magnitude using this model. Additionally, we employed feature importance and principal component analysis to explore the factors influencing proton conductivity. The insights gained from our database and machine learning model are expected to facilitate the targeted design of proton-conductive MOFs.

# INTRODUCTION

As interest in sustainable energy grows, energy storage systems (ESS) such as fuel cells (FC) and batteries are increasingly recognized as promising alternatives to fossil fuels.[1, 2] In particular, proton exchange membrane fuel cells (PEMFC) are noted for their high power density and ease of recharging.[3] The most conventional material for membrane in these cells is the Nafion membrane, which, despite its widespread use, suffers from issues such as low thermal stability, high cost, and low ion selectivity.[4-6] Metal-organic frameworks (MOFs), crystalline porous solid, are emerging as potential materials for these membranes, due to their designability and functionality.[7] Traditionally, MOFs have been utilized in fields such as adsorption[8, 9], separation[10-12], sensing[13, 14], and catalysis[15, 16]. They exhibit high crystallinity, tunable porosity, and the ability to undergo post-synthetic functionalization, demonstrating their potential as solid-state proton conducting membranes.[17, 18] Indeed, proton-conducting MOFs are currently being studied through numerous experiments, yet only a few such materials have been reported.[19] In particular, proton conductivity is influenced by the concentration of mobile protons and mobility by conduction pathway determined by the synergistic effects of various factors such as temperature, humidity, and guest molecules, making it challenging to design a MOF with high proton conductivity.[7, 20] Additionally, while computational simulations aid in analyzing the mechanisms of MOF proton conductivity, obtaining accurate proton conductivity measurements through simulation proves to be difficult and time-consuming.[21, 22]

To address this issue, data-driven research might present a viable solution. Various mining tools and algorithms, including text mining, graph mining, Application Programming Interfaces (APIs), and Generative Pre-Trained Transformers (GPT), have been developed across various materials science fields to obtain specific information about materials.[23-28] Research have been conducted to predict the properties of materials and design materials with specific physical properties by developing machine learning models based on data obtained through these methods. Specifically, Bradford et al. collected data on polymer ionic conductivity from publications to discover solid polymer electrolytes (SPE) with high ionic conductivity.[28] They constructed a model to predict ionic conductivity using the polymer molecular graph and salt molecular graph, polymer molecular weight, salt concentration, and temperature. Notably, the model performed well by incorporating chemical information using the Arrhenius equation.

In addition, transformers have recently demonstrated their efficacy not only in natural language processing (NLP) but also in various materials science fields.[29-35] This involves a pretraining process where general features are learned from extensive datasets, preparing for subsequent specific tasks. Particularly, transfer learning, which leverages these pretrained models and applies them to other datasets, is gaining attention due to its ability to produce relatively high performance quickly.[36] Furthermore, transfer learning with transformers can mitigate the challenge of sub-optimal performance from insufficient data, making it especially advantageous for materials science fields where datasets are typically smaller.[37-40]

In this study, we propose both a descriptor-based machine learning model and a transformer-based model using transfer learning to predict the proton conductivity of Metal-Organic Frameworks (MOFs). We also constructed a database of 248 MOFs, which includes the names of the MOFs, published Digital Object Identifiers (DOIs), proton conductivity values, temperature (T), relative humidity (RH), and guest molecule information. Our models were able to predict the proton conductivity of MOFs with an error margin of approximately one order of magnitude. Given that our approach is based on experimental structures and data, our database and models are capable of reducing trial and error and providing clearer guidance for future experimental work.

# RESULTS AND DISCUSSIONS

## Data Extraction and Curation

First, we compiled a collection of experimental MOF papers relating to the topic of proton conductivity. Utilizing the Python Application Programming Interface (API) for Scopus, we conducted an extensive literature search for articles containing the keywords 'metal AND organic AND frameworks, AND proton AND conductivity', resulting in the identification of Digital Object Identifiers (DOIs) for 741 publications. To obtain the Crystallographic Information Files (CIF) of MOFs, we employed the Python API of the Cambridge Structural Database (CSD)[41], confirming the availability of relevant structural files associated with each DOI. This procedure enabled us to narrow our analysis to 241 articles that had associated Crystallographic Information Files (CIF) and reference codes (**Figure 1a**). The CIF files that were included in the Supporting Information or in previous studies were also incorporated (**Table S1**).

It is well-established that the proton conductivity of MOFs is influenced not just by its structure but also by temperature, relative humidity (RH), and guest molecules. Specifically, proton conductivity generally increases with relative humidity (except in anhydrous conditions), temperature, and the presence of guest molecules.[7] Data points indicating a significant decrease in proton conductivity at high temperatures due to the breaking hydrogen bond by loss of guest molecules, like water and dimethyl ammonium cations, were excluded from our analysis.[42] In the majority of the experimental papers, proton conductivity data is graphically represented against temperature or relative humidity (RH), as illustrated in **Figure 1a**. A digitizer was employed to extract data from these graphs. The compiled data encompasses DOI, Name (CSD reference code), proton conductivity, temperature, RH, and guest molecules (**Figure 1a**).

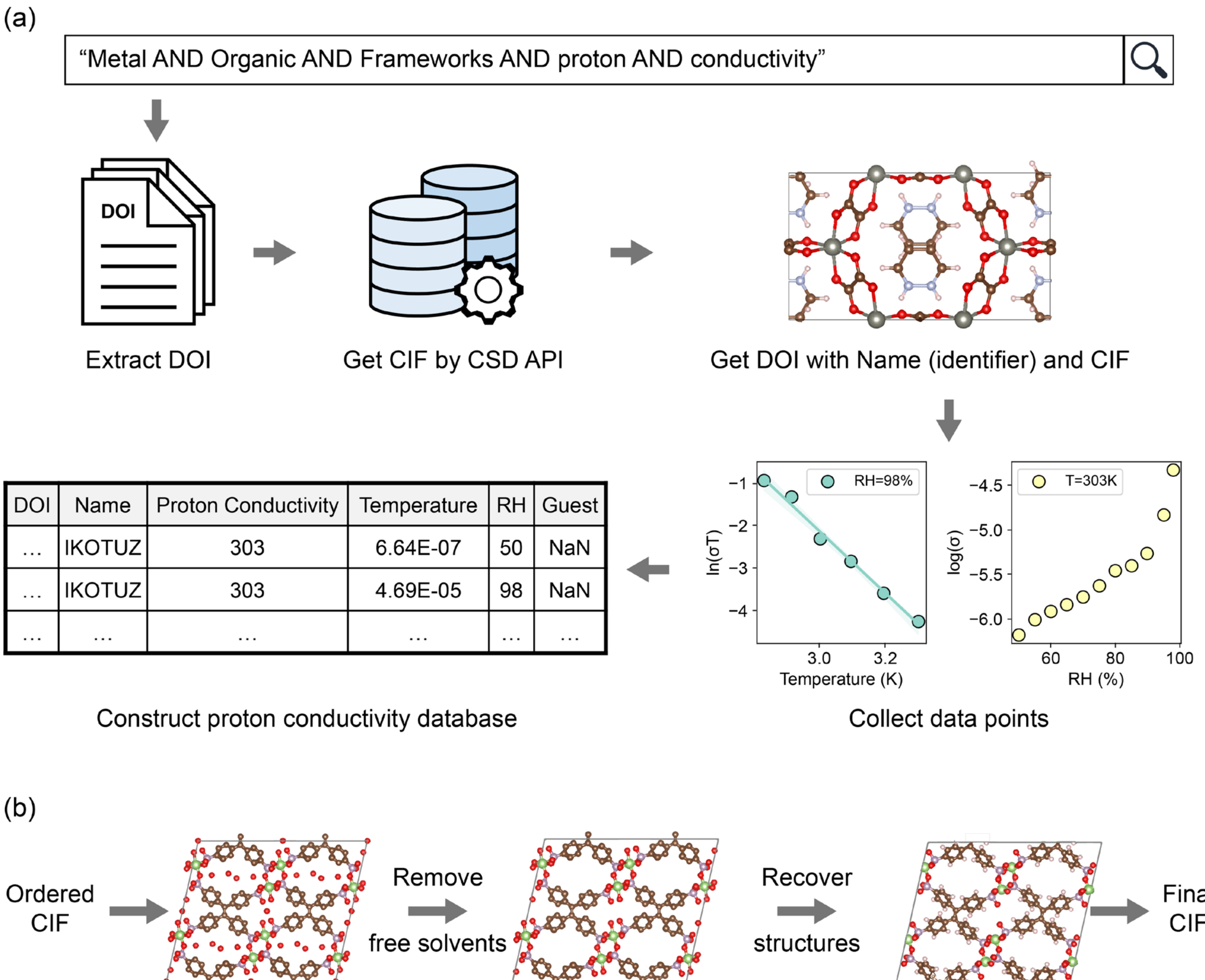

**Figure 1** Data Collection Process for Overall Proton Conductivity. (a) Illustration of the process of building a proton conductivity database. (b) Details of the process of refining the experimental structures of Metal-Organic Frameworks (MOFs), focusing specifically on the purification process of the ordered structures.

After extracting the data, we employed a refinement process akin to that used in the CoRE MOF database[43] for the experimental structures corresponding to the CSD reference codes, detailed in **Figure 1b**. Initially, any crystal structure with inter-atomic distances less than 0.6 Å was categorized as disordered, while the remainder were considered ordered. From the ordered structures, we eliminated free solvent molecules. Subsequently, we utilized mofchecker[44] to identify structures with overcoordinated or undercoordinated C, N, and H atoms. Subsequently, we recovered and optimized these structures using Materials Studio[45]. Specifically, for the structure shown in Figure 1(b), H atoms were added to the undercoordinated C atoms and then optimized. Furthermore, we directly recovered disordered structures, while those deemed challenging to correct were disregarded (**Figure S2**). Through this meticulous refinement procedure, we developed a database for the proton conductivity of MOFs with a total of 3,388 data points, 172 DOIs and 248 CIF files. Publisher information for each DOI journal is shown in **Table S2**. The distribution of these data points according to proton conductivity, temperature, RH, and types of guest molecules is illustrated in **Figure 2**.

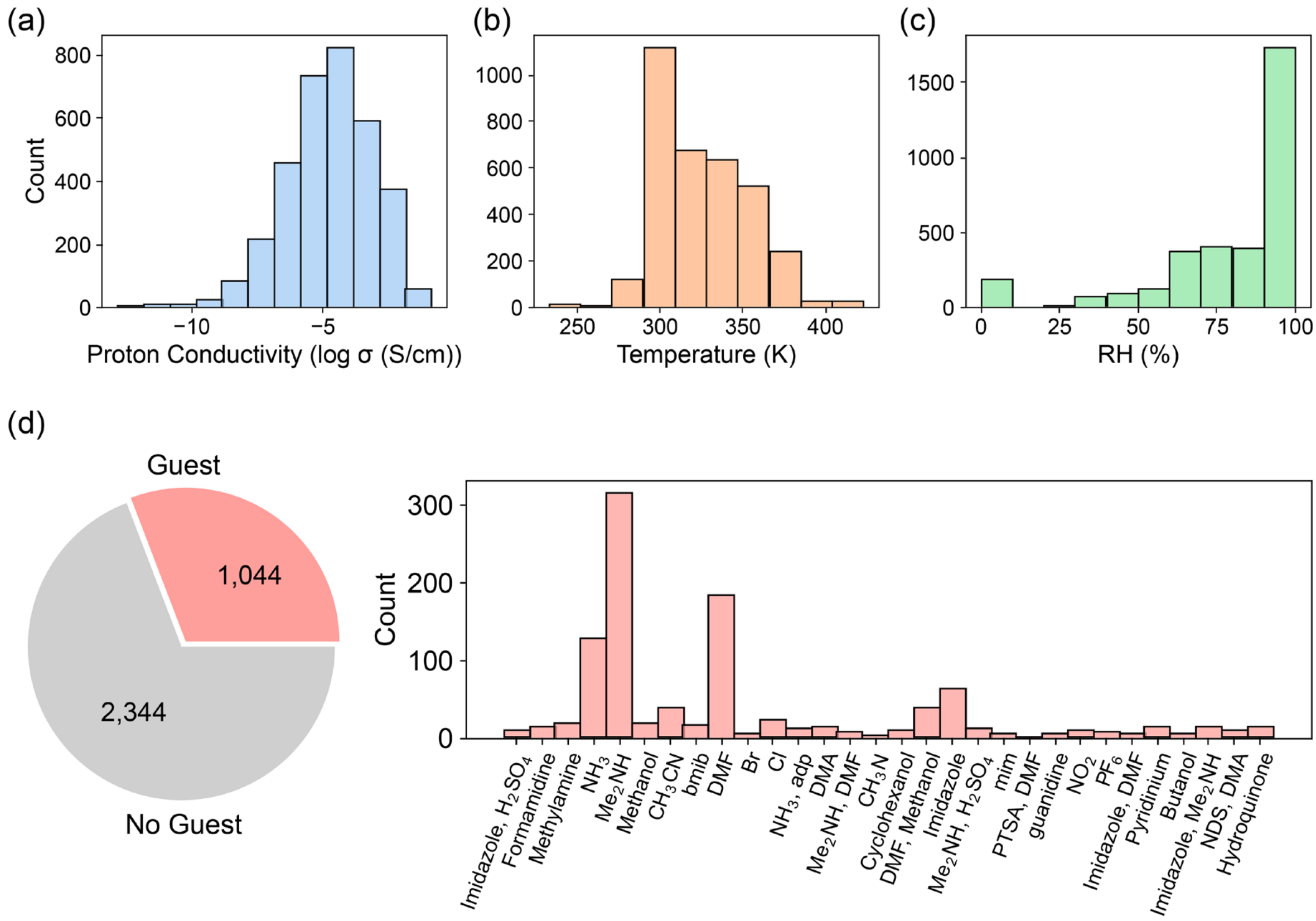

**Figure 2** Distribution of Collected Data on Proton Conductivity, Temperature, Relative Humidity (RH), and Guest Contributions. This figure shows how proton conductivity varies with temperature, RH, and different guests in the dataset.

## Machine Learning Model Construction

In this work, proton conductivity of MOFs were predicted using two approaches. The first method involved extracting descriptors from the MOFs and the guest molecules to predict proton conductivity through a machine learning model (**Figure 3a**). The second approach utilized transfer learning with a Transformer model (**Figure 3b)**.

For the machine learning model, MOF descriptors were derived by extracting revised autocorrelations (RACs) and geometric features from CIF files converted to a P1 structure. RACs are graph-based descriptors reflecting products and differences of five heuristic atom-wise properties: nuclear charge (Z), topology (T), identity (I), covalent radius (S), and electronegativity ($\chi$). These RACs were acquired through MolSimplify's code, encompassing 160 descriptors.[26, 46] A total of 14 geometric features, including pore size, volume, and surface area, were computed using the Zeo++ code using a probe with a radius of 1.2 Å.[47] Detailed information on each feature is provided in **Table S2**. Guest molecules were sourced from PubChem, converted to Mol format through RDKit module, and subsequently, two-dimensional features were extracted using MoleculeDescriptors of RDKit, resulting in 199 features.[48] In total, we assembled a dataset of 375 features, incorporating temperature and relative humidity. The machine learning model employed comprised Artificial Neural Networks (ANN), Gaussian Process Regression (GPR), and XGBoost.

For transfer learning, we utilized two pretrained transformer models: the MOFTransformer[35] and ChemBERTa[30]. The MOFTransformer is designed to account for both local and global features through atom-based graph embeddings and energy grid embeddings, enabling it to predict various properties across different MOF structures. Atom-based graph and energy-grid embeddings were generated from each MOF's CIF file for input of the MOF Transformer. ChemBERTa is a pretrained model for the Simplified Molecular Input Line Entry System (SMILES)[49] of 77 million chemical structures from PubChem, which is suitable for understanding guest molecules obtained from PubChem. The structure of these guest molecules was converted into canonical SMILES format using the RDKit module. We utilized the CLS token from each pretrained transformer model, both of which have 768 dimensions. Temperature (T) and relative humidity (RH) were separately embedded and then combined through element-wise addition with the CLS token from each transformer. Element-wise addition involves combining two characteristic vectors by adding their corresponding elements. This combined data was then processed through a neural network to build our model, as illustrated in **Figure 3b**. In transfer learning, there exist various approaches, including retraining (or fine-tuning) the entire model to adapt to the dataset, which involves unfreezing all layers for training, and training only specific layers to be updated while some layers remain frozen.[37] In our study, we explored two distinct methods: one where all layers of the transformer models were trained (unfreezing and finetuning) and the other where the layers up to the CLS token were frozen, preventing them from being updated during training (freeze). Detailed information on the model structure and parameters is provided in the **Methods** section.

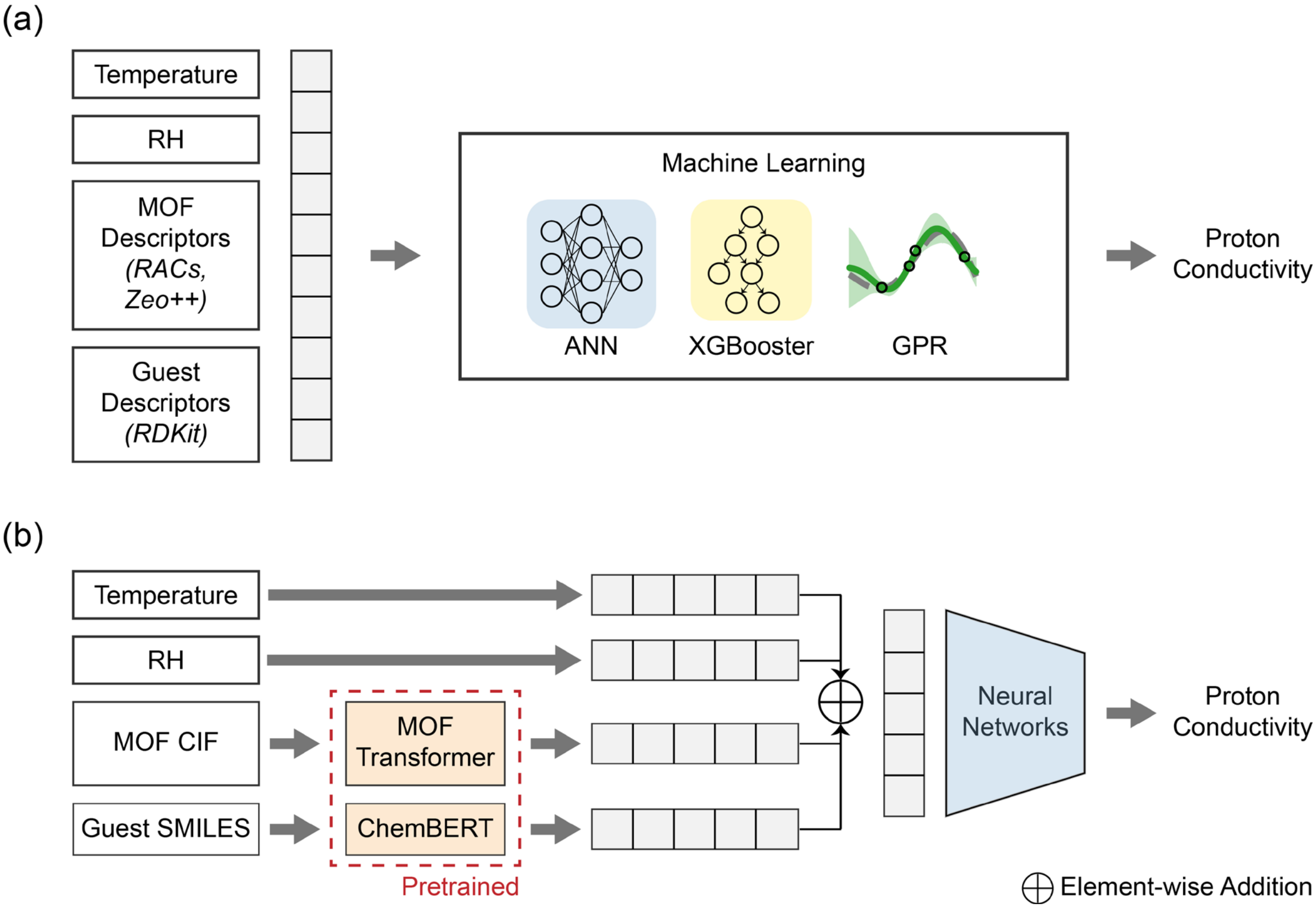

**Figure 3** Schematic Diagram of Models for Predicting Proton Conductivity. (a) Descriptor-based machine learning model using descriptors of Metal-Organic Frameworks (MOFs) and guest molecules. (b) Transfer learning model that employs pre-trained Transformers for Metal-Organic Frameworks (MOFs) and guest molecules.

## Performance Comparison

To train the models, we divided the dataset into train, validation, and test sets and ensured that the same MOF does not appear more than once in any of these sets. This ensured that MOFs seen during the training were not used for validation or test set, which would immensely facilitate the prediction. Having the same MOF in both the training and the other sets (with different conitions) lead to extremely high-performance results (**Figure S3**), but such a split may lead to biased predictions for MOFs that were already introduced during training when validation or testing proceeds. Consequently, we split the training and test sets to 0.8/0.2 ratio, based on MOF structure. Furthermore, the training dataset was further divided into training and validation sets using a 5-fold cross-validation process, ensuring rigorous evaluation and optimization of the prediction capabilities.

**Table 1** Performance Comparison of Descriptor-Based Model and Transformer-Based Model. Performance was evaluated using Mean Absolute Error (MAE) of the test dataset, derived from 5-fold cross-validation.

| **Model** | | **Test MAE (log (S/cm))** |
|---|---|---|
| Descriptor-based | Descriptors + ANN | $1.22 \pm 0.07$ |
| | Descriptors + GPR | $1.2 \pm 0.04$ |
| | Descriptors + XGBoost | $0.98 \pm 0.07$ |
| Transformer-based | Transfer Learning (Freeze) | $0.91 \pm 0.04$ |
| | Transfer Learning (Unfreeze, finetuning) | $0.98 \pm 0.05$ |

The performance of the model is represented by the Mean Absolute Error (MAE), as detailed in **Table 1** (a Mean Absolute Error of one indicates that the proton conductivity deviates from the actual value by about one order of magnitude on average). In terms of performance, XGBoost exhibited the best results among the descriptor-based models, while transfer learning (freeze) outperformed all models, achieving the highest performance. Transfer learning, leveraging a transformer pretrained on general features of MOFs and guest molecules, demonstrates superior performance compared to other models, enhancing prediction accuracy for specific tasks. Within the transfer models, the freeze approach surpasses the unfreeze (finetuning) method in performance. It is attributed to the small size of our dataset, where training all layers of the transformer (unfreeze, finetuned) could lead to overfitting. Moreover, freezing certain layers not only boosts performance but also expedites the training process due to the focused training on specific layers.

In our transfer learning approach, we integrated the CLS tokens of MOFs, CLS tokens of guests, temperature embedding vectors, and RH embedding vectors through element-wise addition. We also utilized the methods involving concatenation and the Arrhenius equation[28] for transfer learning. The concatenation model increases the feature space by stacking the feature vectors of the MOF, the guest molecule, temperature, and RH, while the Arrhenius equation model is a temperature-activated process that stacks the feature vectors of the MOF, the guest, and RH. The use of element-wise addition yielded better results compared to either the application of concatenation or the Arrhenius equation (**Figure 4 and Figure S4**). Contrary to previous findings, the performance of the model using the Arrhenius equation was lower than that using concatenation alone, possibly due to the complex structure of MOFs compared to polymers and the scarcity of data. Indeed, the training performance of models employing the Arrhenius equation was also lower than that of other models. Simpler approaches may be more appropriate when data is scarce and complex.

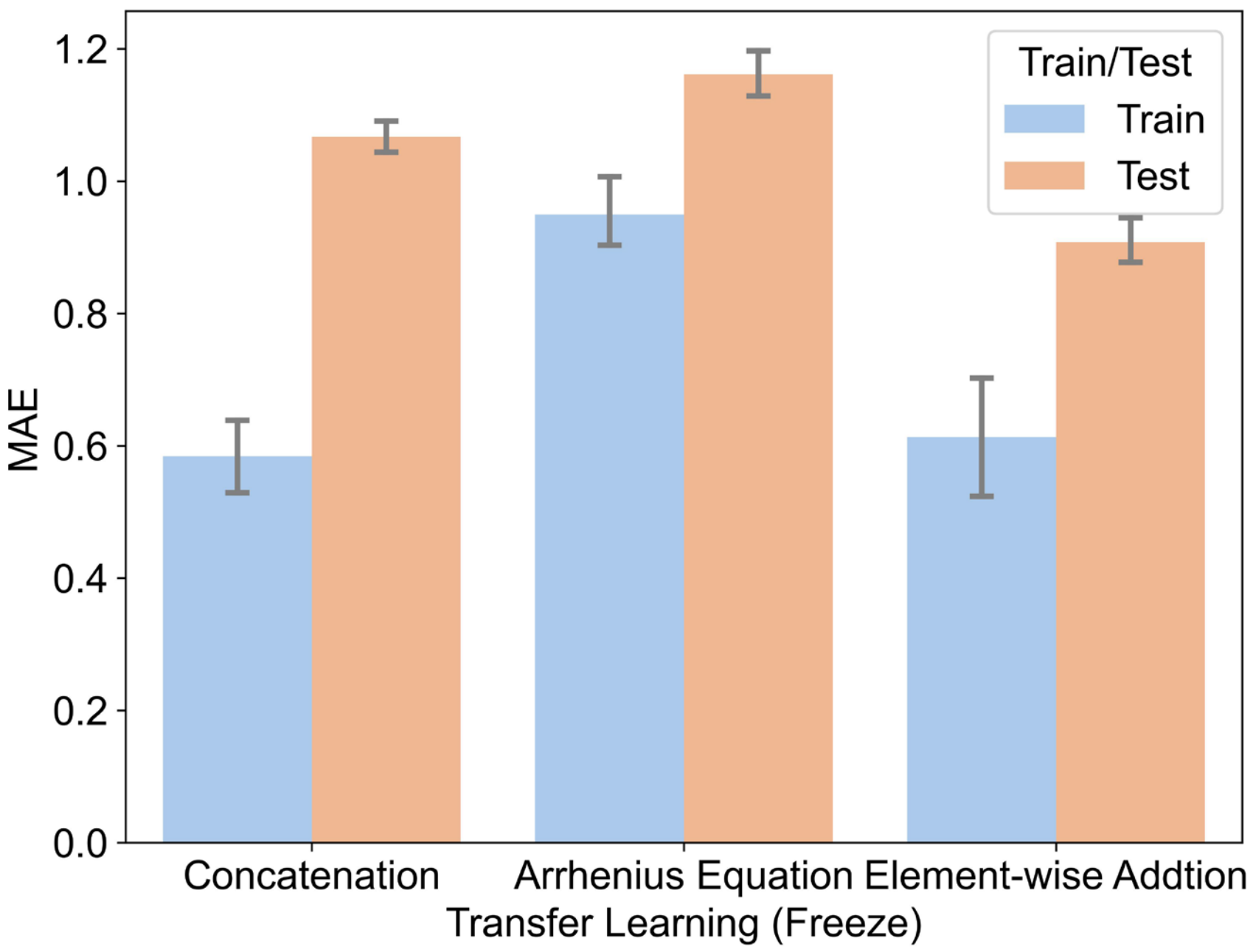

**Figure 4** Comparison of model performance using different methods for combining the information (T, RH, Guest, MOFs) in the Transfer Learning (Freeze) Model. Performance is assessed through the Mean Absolute Error (MAE) from 5-fold cross-validation on the test dataset.

We conducted an in-depth analysis of the XGBoost model and transfer learning (freeze), which demonstrated superior performance among various machine learning models. We focused on the importance of descriptors in predicting proton conductivity for the XGBoost model, specifically assessing feature importance (**Figure S5**). While the important features varied slightly across different cross-validation processes, descriptors of guest molecules consistently appeared to be the most important, followed by differences in linker connections of MOFs. Given that our dataset includes numerous data points with variations in temperature, humidity, or guest molecules for identical MOFs, we also meticulously assessed the effects of each of these factors. Furthermore, we compared the performance variation when training XGBoost with different combinations of descriptors: temperature and humidity (T, RH), temperature, humidity, and guest (T, RH, guest), and temperature, humidity, guest, and MOF (T, RH, guest, MOF) on the same datasets. The results indicated that while proton conductivity could initially be predicted with only temperature and humidity, incorporating guest and MOF information significantly enhanced the performance of the model (**Figure 5a to c**). This finding highlights the importance of temperature and humidity as fundamental factors in predicting proton conductivity but also emphasizes the crucial role of guest and MOF information for achieving higher accuracy. In line with this, our analysis of the transfer learning (freeze) model using Principal Component Analysis (PCA) revealed that humidity and temperature are the principal components (**Figure 5d and 5e**). Specifically, principal component 1 was mapped to relative humidity (RH) and principal component 2 to temperature. Notably, the points when the humidity was 0 and when it was greater than 0 were clustered separately. This indicates that lower humidity generally corresponds to lower proton conductivity, while in an anhydrous system, proton conductivity behaves differently.

Thus, we have used the structural information of the MOF and measurement conditions such as guest, relative humidity (RH), and temperature to predict the proton conductivity of MOFs. We have developed a machine learning model predicting proton conductivity with a mean absolute error (MAE) of about 0.91. In the model analysis, guest descriptors were identified as an important factor in predicting proton conductivity. We anticipate that the model's performance could be slightly enhanced if information on the quantity of guests, particularly water adsorption, were available. Furthermore, while coordinated water is known to affect proton conductivity, in some experimental structures all solvents had already been removed, making it impossible to determine the amount of coordinated water. This was assessed by analyzing open metal sites (OMS) and calculating the number of water molecules coordinated per metal atom, as shown in **Figures 5g** and **5h**. In MOFs with open metal sites, OMS should disappear as water coordinates, but in structures where solvents have already been removed, OMS may appear uncoordinated with water. Accurately determining the amount of water per metal in the MOF could potentially enhance the prediction performance of proton conductivity.

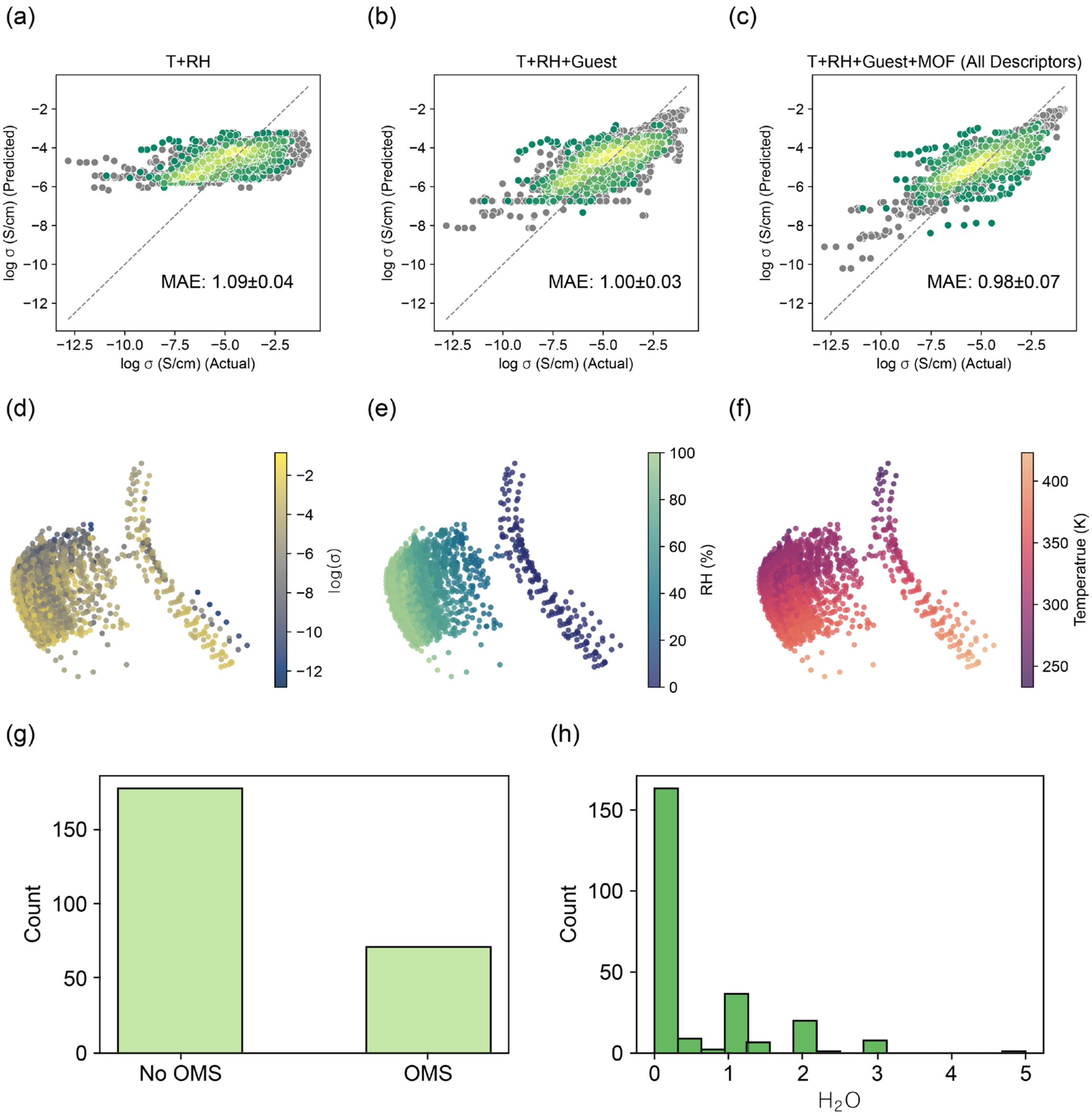

**Figure 5** Analysis of the XGBoost Model and the Transfer Learning (Freeze) Model. (a)–(c) Performance comparison of different descriptor combinations in the XGBoost Model, evaluated using Mean Absolute Error (MAE) from 5-fold cross-validation of the test dataset. (d)–(f) Dimensionally reduced scatter plots through Principal Component Analysis (PCA) for the Transfer Learning (Freeze) Model. Each point represents data (T, RH, Guest, MOF), categorized by (d) proton conductivity, (e) RH, and (f) temperature. (g) Histogram showing whether MOFs have open metal sites (OMS) (h) Histogram depicting the number of water molecules coordinated per metal in MOF structures.

.

# CONCLUSIONS

In this work, we gathered experimental data on the proton conductivity of metal-organic frameworks (MOFs) and employed machine learning techniques to obtain the structure-property relationship. We compiled an experimental database incorporating varying temperatures, relative humidity, and guest molecule information with 248 MOF structural files. We developed machine learning models to predict the proton conductivity of MOFs using both descriptor-based and transformer-based approaches. Notably, the transformer-based transfer learning model (Freeze) demonstrated superior performance, achieving a Mean Absolute Error (MAE) of 0.91. Further, our feature importance analysis for XGBoost model revealed that variations in MOF linker connections significantly influence proton conductivity. We also investigated the effects of temperature and humidity on performance by using various combinations of descriptors in the XGBoost model and PCA in the transfer learning model. Given that our model integrates experimental structures and data, we anticipate that our work will serve as a valuable tool in experimentally identifying MOFs with optimal proton conductivity.

# METHODS

*Descriptor-based model*

The descriptor-based model employed three different models: Artificial Neural Networks (ANN), Gaussian Process Regression (GPR), and XGBoost. In total, there exist 375 descriptors: 174 for the MOF, 199 for the guest, and descriptors for temperature (T) and relative humidity (RH). When more than one guest was present, the guest descriptor was averaged. The ANN configuration includes two hidden layers each with 400 dimension. Batch normalization and dropout with a rate of 0.5 were implemented. The Sigmoid Linear Unit (SiLU) was used as the activation function. The training process consisted of 30 epochs with a warm-up ratio of 0.05 and a batch size of 128. AdamW was employed as the optimizer with a weight decay of 0.01, utilizing the Noam scheduler with an initial rate of 5e-5, a maximum rate of 1e-4, and a final rate of 1e-5. For GPR, we used Gpytorch's ExactGP model with a Radial Basis Function (RBF) Kernel, a learning rate of 0.1, and the Adam optimizer. XGBoost employed a tree-based model with a learning rate of 0.3. All models incorporated early stopping with a patience setting of 10. Descriptors and proton conductivity values in all models were standardized using a standard scaler.

*Transformer-based model*

The Transformer-based model employed transfer learning techniques, including freezing (where the layers of the MOF-transformer and ChemBERT are fixed) and unfreezing with finetuning (where all layers are trained together). The CLS token for both the MOF transformer and ChemBERT has 768 dimensions and is processed through a dense layer maintaining 768 dimensions. The embedding layers for temperature (T) and relative humidity (RH) were also configured to 768 dimensions. In cases where two guests were present, the ChemBERT CLS tokens for each guest were averaged. The embedding vectors for T, RH, MOF, and Guest were combined element-wise and then passed through two hidden layers, each with 400 dimensions. AdamW was utilized as the optimizer, accompanied by a weight decay of 0.01, and the Noam learning rate scheduler was employed. The initial learning rate was set at 1e-5, the maximum at 5e-5, and the final at 1e-5. The training consisted of 30 epochs with a warm-up ratio of 0.05. Batch normalization, a dropout rate of 0.5, and the Sigmoid Linear Unit (SiLU) activation function were applied. The batch size for the transfer learning (Freeze) model was set at 128, while for the transfer learning (Unfreeze, finetuning) model, it was reduced to 32. Early stopping was implemented with a patience of 10 and 5, respectively. Additionally, temperature, humidity, and proton conductivity in all models were normalized using a standard scaler.

## Code availability

The code is available at https://github.com/seunghhs/ProtonMOF.git.

## ASSOCIATED CONTENT

**Supporting Information**. The following contents are included: The CSD Identifiers for CIF in the Supporting Information or in the. Distribution of cleaned data collected. Examples of direct recovery of disordered structures. Publisher information. The detail information about MOF descriptors. Training results using a randomly divided dataset. Scatter plot of model performance, Feature importance of the XGBoost model

## AUTHOR INFORMATION

### Corresponding Author

* Email: jihankim@kaist.ac.kr

Author Contributions

S.H. and J.K. conceived the idea, planned modeling, implemented the work with the assistance of B.L. and D.L. S.H. and J.K wrote the manuscript with inputs from all the authors. All authors contributed to discussions informing the research.

**ORCID**

Seunghee Han: 0000-0001-8696-6823

Byeong Gwan Lee: 0009-0008-9360-7771

Dae-Woon Lim: 0000-0002-6389-4596

Jihan Kim: 0000-0002-3844-8789

**Notes**

The authors declare no competing interest.

ACKNOWLEDGEMENT

This work was supported by the National Research Foundation of Korea (NRF) (RS-2024-00337004) and by the National Supercomputing Center with supercomputing resources including technical support (KSC-2024-CRE-0189)

# Supporting Information

# Machine Learning Based Prediction of Proton Conductivity in Metal-Organic Frameworks

*Seunghee Han[1], Byoung Gwan Lee[2], Dae Woon Lim[2], and Jihan Kim[1*]*

1 Department of Chemical and Biomolecular Engineering, Korea Advanced Institute of Science and Technology, Daejeon 34141, Republic of Korea.

2 Department of Chemistry and Medical Chemistry, Yonsei University, Wonju, Gangwondo 26493, Republic of Korea

*Corresponding author: jihankim@kaist.ac.kr

**Table S2** CSD Identifiers (or Names) and DOIs for CIF included in the Supporting Information or in previous research

| DOI | CSD Identifier (or Name) |
|---|---|
| 10.1002/smll.202301122 | KOKKOL, MIL-88B-LB, MIL-88B-SB, KUXREC |
| 10.1021/acs.inorgchem.1c01522 | FATKIW, FATKAO, FATKOC, KESHAT, FATKUI, FATKES |
| 10.1021/acs.jpcc.6b04649 | WUTBEU |
| 10.1021/acsaem.2c03518 | RUBTAK, CPO-27-Mg-NCSMA, CPO-27-Mg-NCSFA, CPO-27-Zn-NCS, CPO-27-Ni-NCS, CPO-27-Mg-NCS |
| 10.1021/jacs.8b06582 | KAUST7 |

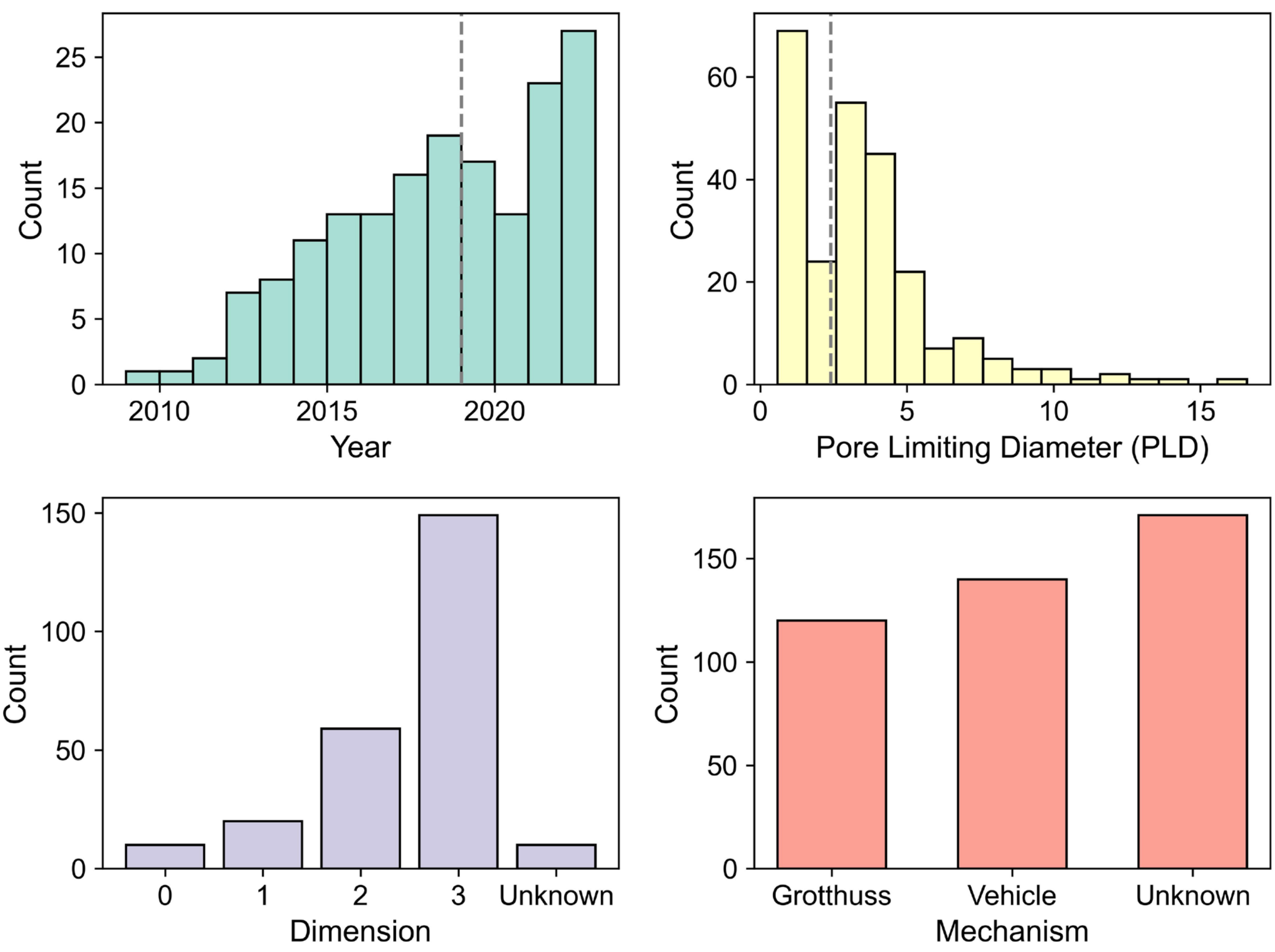

**Figure S6** Distribution of Cleaned Data Collected, Categorized by: (a) Year of publication for 172 DOIs (b) Pore Limiting Diameter (PLD), calculated using Zeo++ code for 248 MOF structures (c) Dimension, determined by the variance between the existing degree of connection and that of an expanded 2x2x2 supercell for 248 MOF structures; (d) Mechanism, describing the distribution of 431 humidity data for 248 MOF structures. The Activation energy ($\boldsymbol{E_a}$) is categorized as Unknown if not listed, Grotthuss if $\boldsymbol{E_a} < 0.4$, and Vehicle if $\boldsymbol{E_a} > 0.4$.

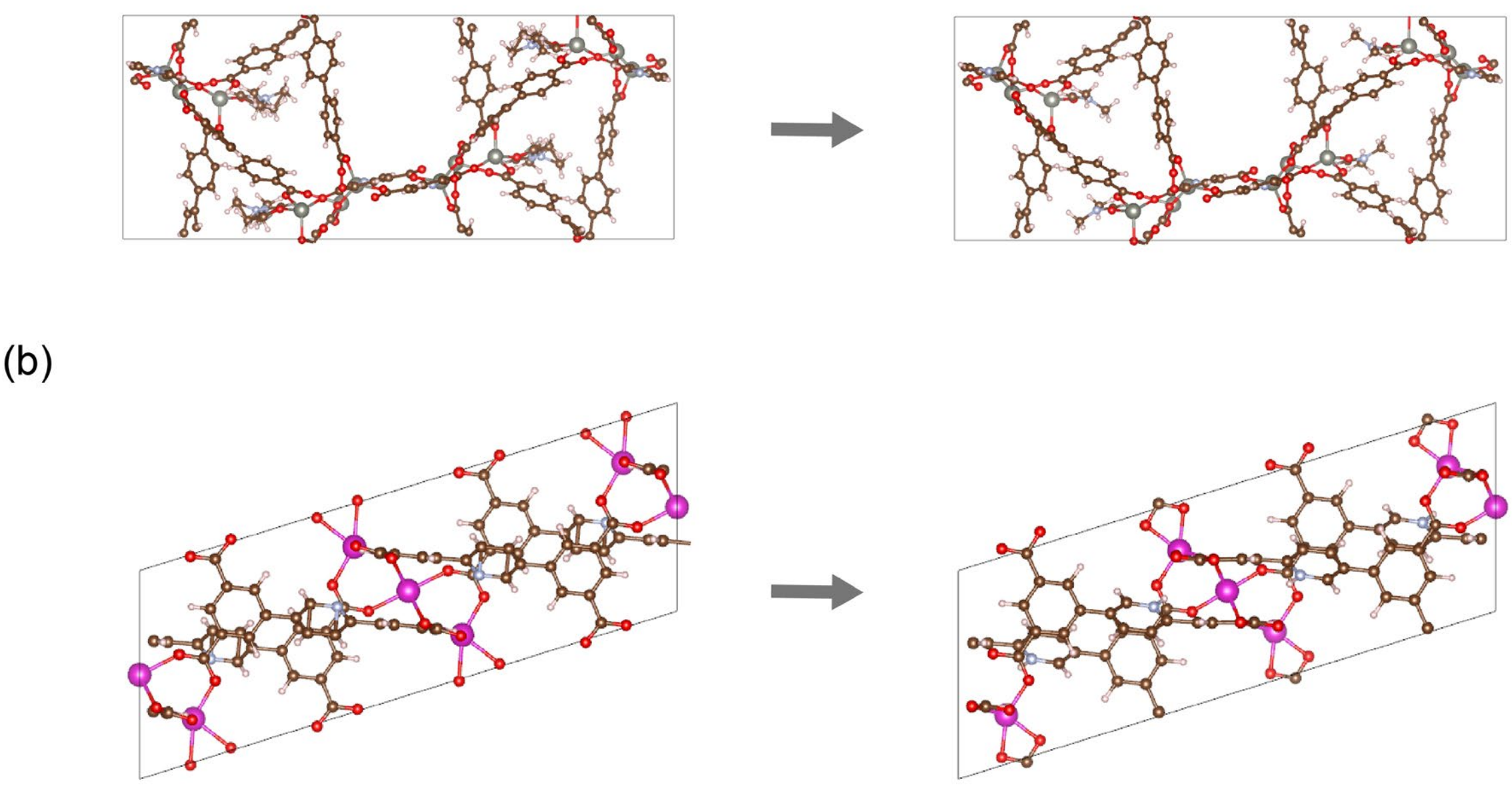

**Figure S7** Examples of direct recovery of disordered structures. (a) MAZYUK: Recovery of the disordered solvent, Dimethylformamide (DMF). (b) AQOREG: Removal of duplicate and disordered atoms in the linker.

**Table S3** Publisher information for 172 DOIs.

| Publisher | Prefix | Count |
|---|---|---|
| RSC | 10.1039 | 66 |
| ACS | 10.1021 | 63 |
| Wiley | 10.1002 | 21 |
| Elsevier | 10.1016 | 17 |
| Taylor and Francis | 10.1080 | 2 |
| Nature | 10.1038 | 1 |
| IUCr | 10.1107 | 1 |
| Chinese Journal of Structural Chemistry | 10.14102 | 1 |

**Table S4** The detail information about MOF descriptors: RACs descriptors and Geometric descriptors. Geometric descriptors are followed: included sphere ($\mathbf{D_i}$), free sphere ($\mathbf{D_f}$), included sphere along the free sphere path ($\mathbf{D_{if}}$), gravimetric pore-accessible volume (GPOAV), gravimetric pore non-accessible volume (GPONAV), gravimetric pore volume (GPOV), gravimetric surface area (GSA), pore-accessible volume (POAV), pore accessible void fraction (POAV_vol_frac), pore non-accessible volume (PONAV), pore non-accessible void fraction (PONAV_vol_frac), volumetric pore volume (VPOV), volumetric surface are (VSA), and density (ρ).

| | Scope | quantity | Count |
|---|---|---|---|
| RAC Descriptors | Metal centered (mc, $D_{mc}$) | Nuclear charge (Z), topology (T), Identity (I), covalent radius (S), electronegativity (χ) | 20 product, 20 difference |
| | Linker connected (lc, $D_{lc}$) | | |
| | Functional group (func, $D_{func}$) | | |
| | Full unit cell (f) | | 20 product |
| | Full linker (f-lig) | | |
| Geometric Descriptors | Geometry | $D_i, D_f, D_{if}$, GPOAV, GPOVAV, GPOV, GSA, POAV, POAV_vol_frac, PONAV, PONAV_vol_frac, VPOV, VSA, ρ | 14 features |
| Sum | | | 174 |

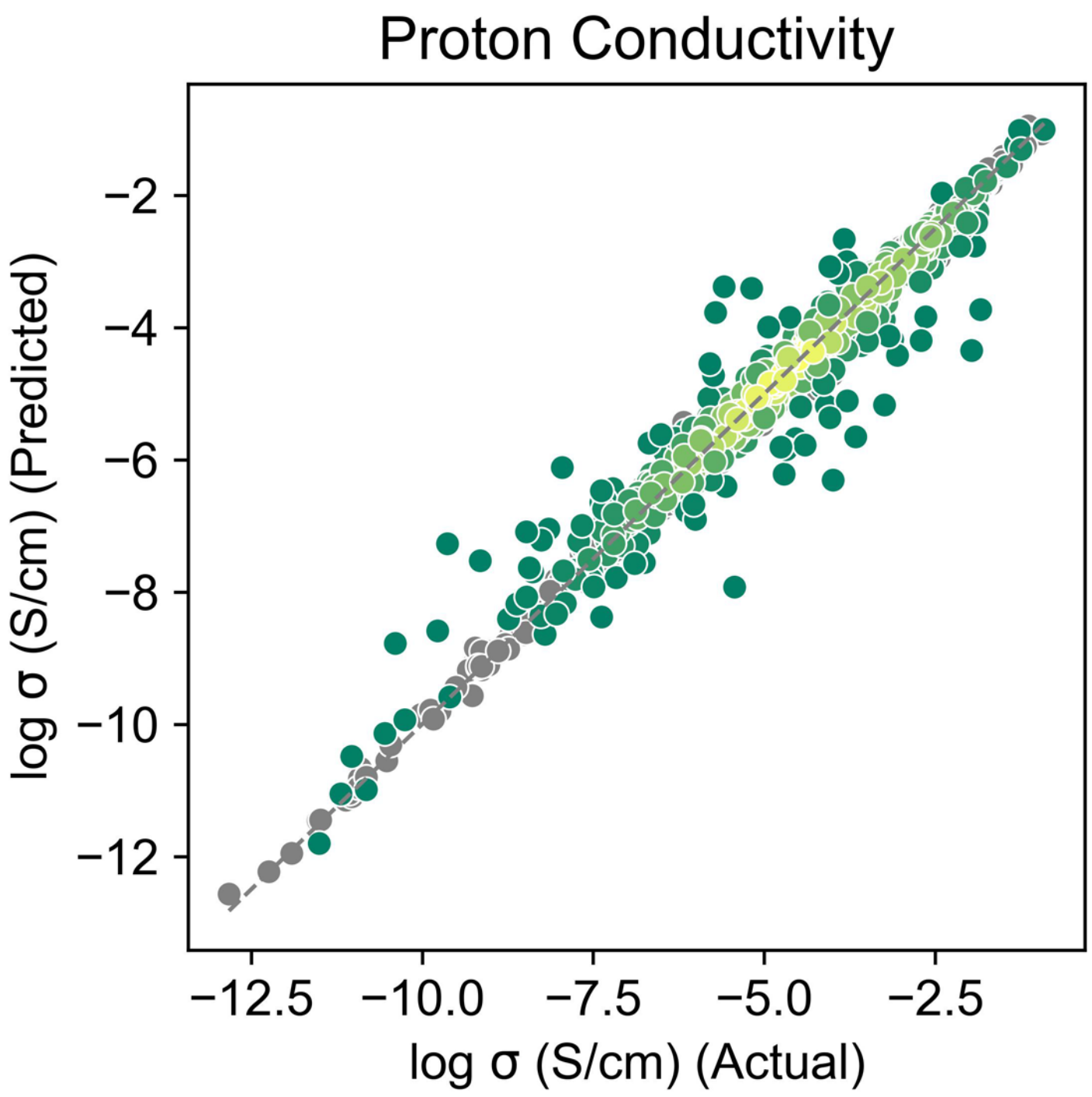

**Figure S8** Training results of the proton conductivity model (Descriptors + XGBoost) using a randomly divided dataset (Train/Validation/Test Split: 0.64/0.16/0.2). Each point represents a data point (T, RH, Guest, MOF). The training dataset is depicted in gray, and the test dataset in green, with color brightness indicating data density. The validation dataset is not shown. The Mean Absolute Error (MAE) values are 0.08 for the training dataset and 0.28 for both the validation and test datasets.

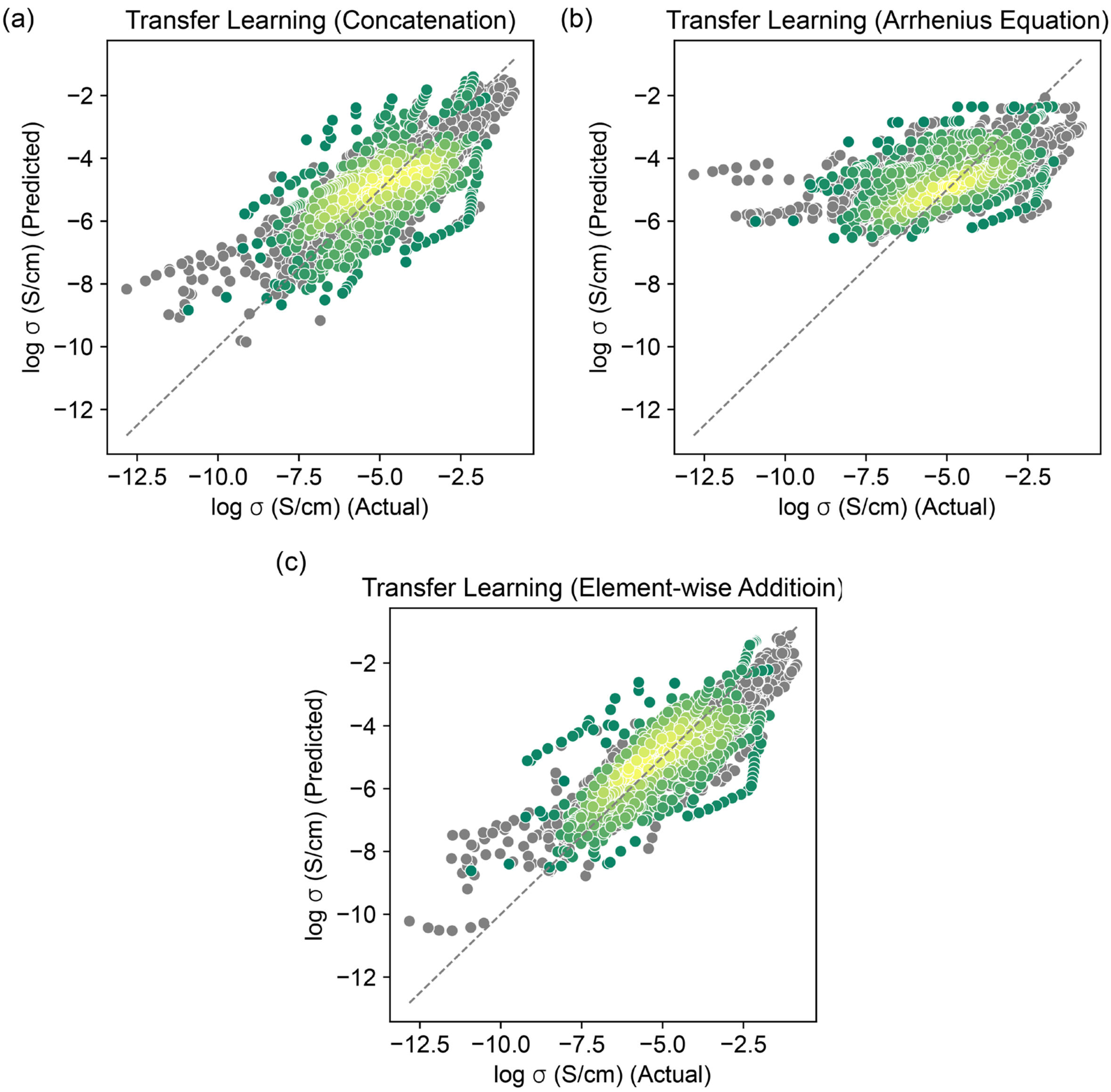

**Figure S9** Scatter plot of model performance. Each point on the plot represents a data entry (T, RH, Guest, MOF). The training dataset is depicted in gray, and the test dataset in green, with the brightness of the color indicating the density of points. The validation dataset is omitted.

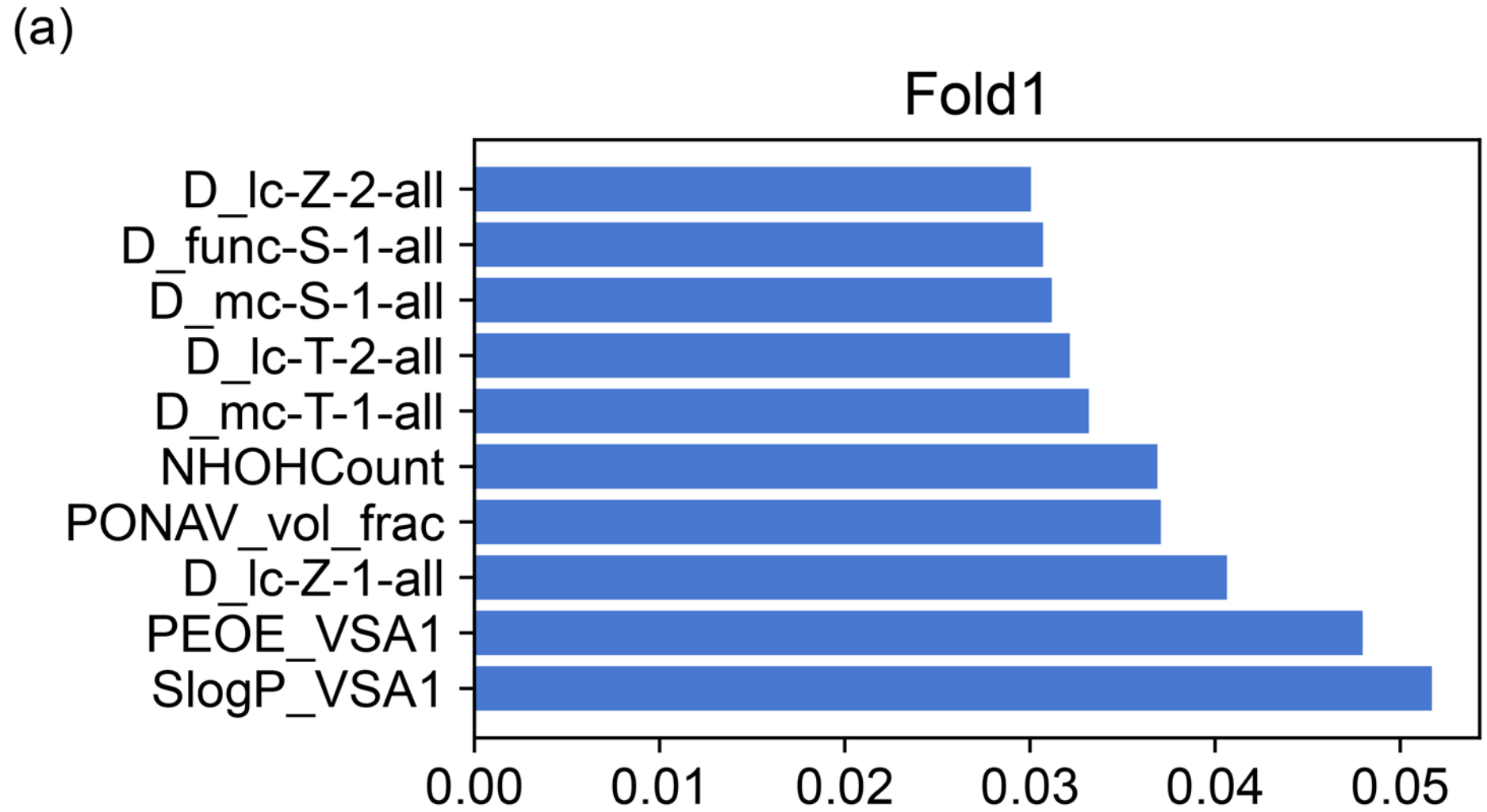

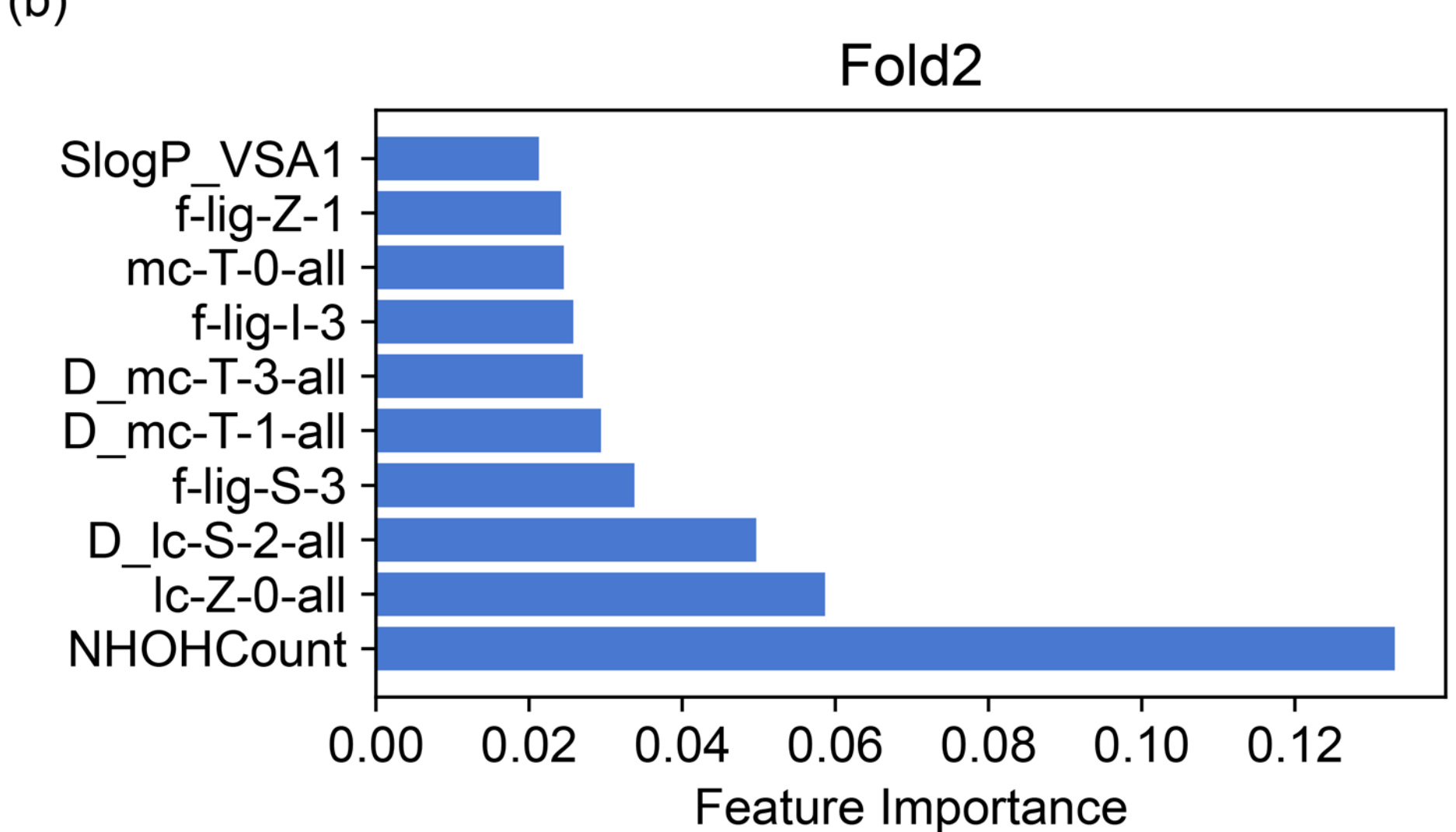

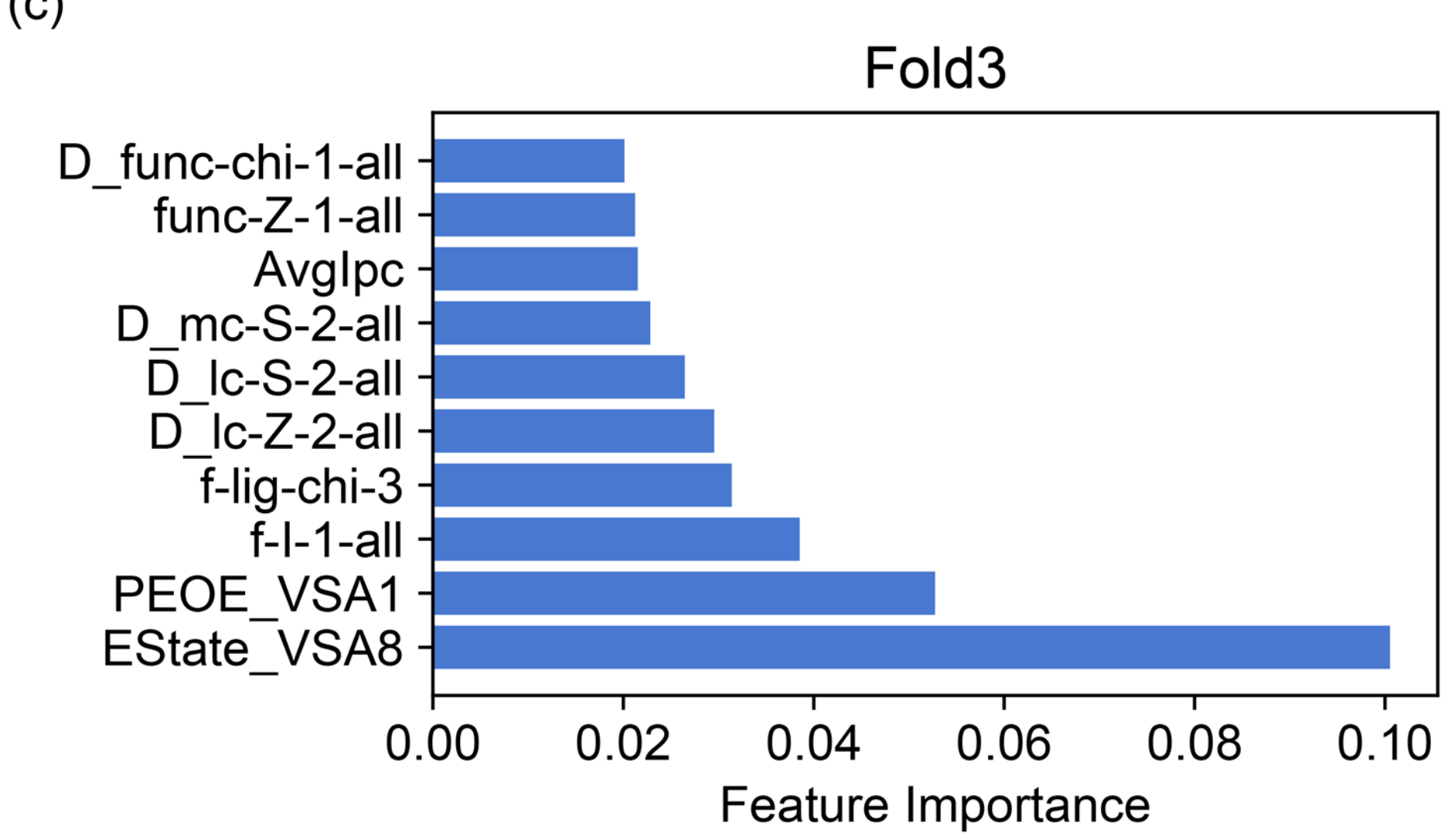

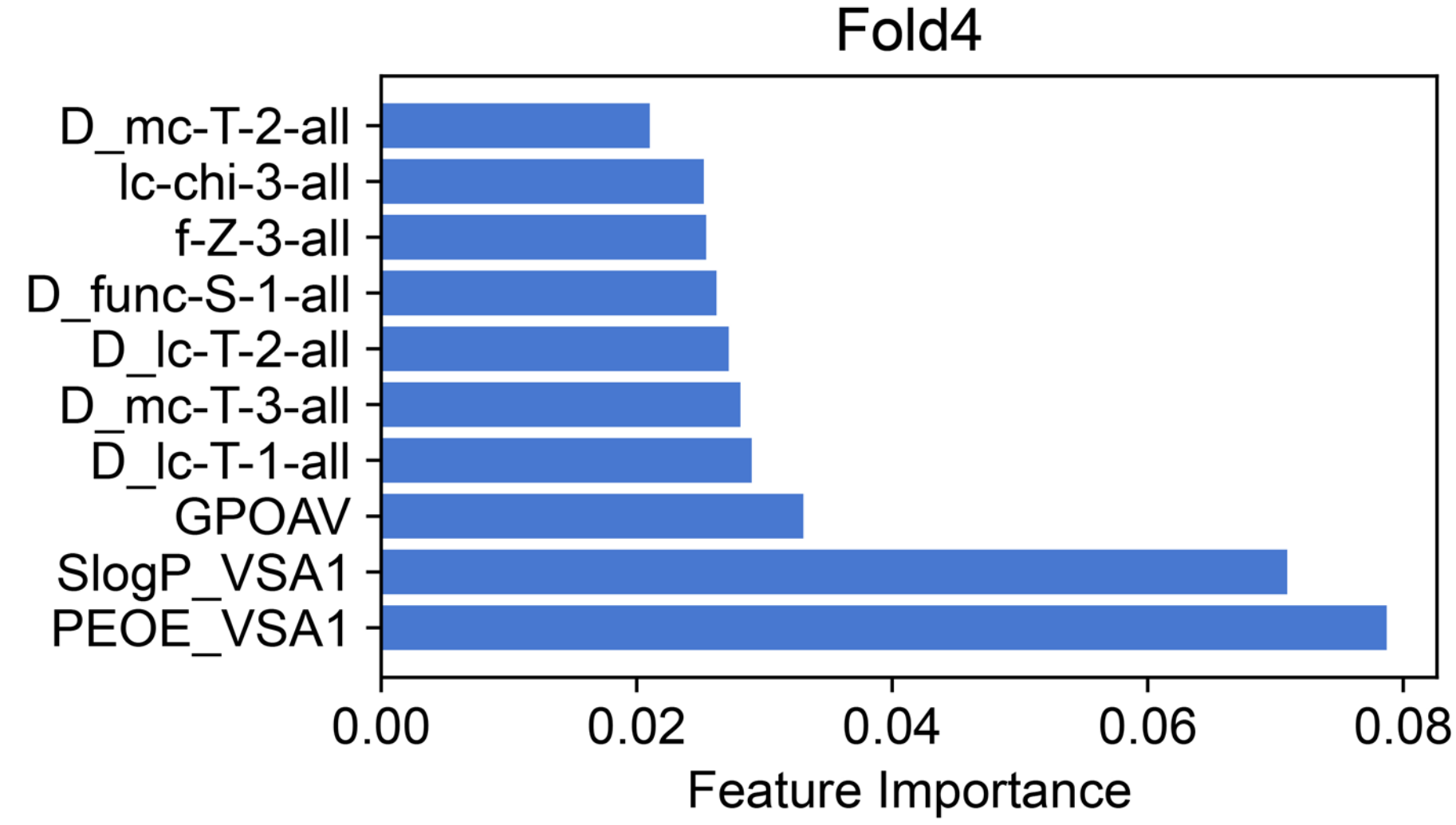

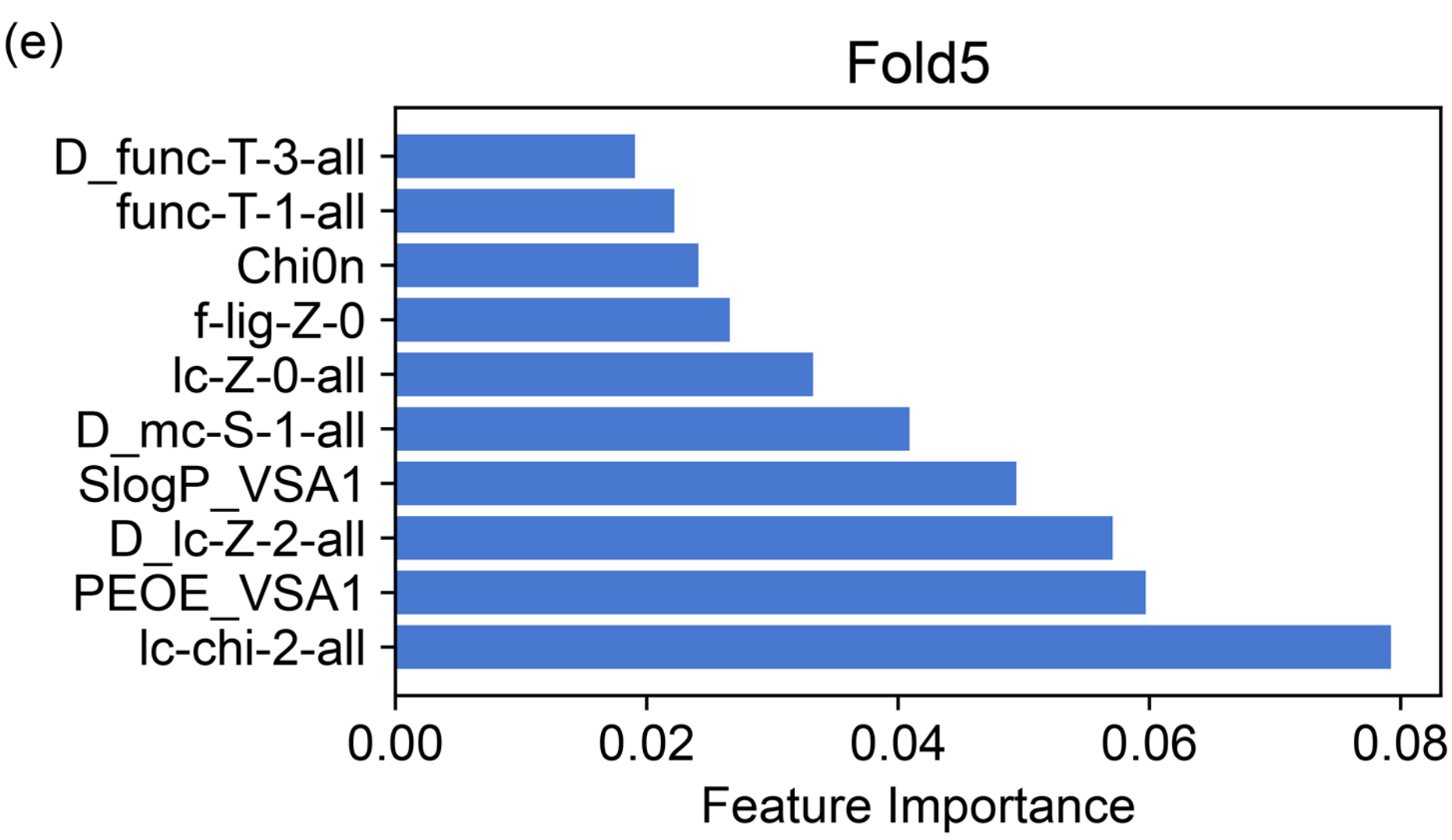

**Figure S10** Feature importance for each fold of the XGBoost model. Only the top 10 most important features are shown for each model fold.